# Multi-Scaling Allometric Analysis for Urban and Regional Development

Yanguang Chen

(Department of Geography, College of Environmental Sciences, Peking University, Beijing 100871, PRC. Email: chenyg@pku.edu.cn)

**Abstract:** The concept of allometric growth is based on scaling relations, and it has been applied to urban and regional analysis for a long time. However, most allometric analyses were devoted to the single proportional relation between two elements of a geographical system. Few researches focus on the allometric scaling of multielements. In this paper, a process of multiscaling allometric analysis is developed for the studies on spatio-temporal evolution of complex systems. By means of linear algebra, general system theory, and by analogy with the analytical hierarchy process, the concepts of allometric growth can be integrated with the ideas from fractal dimension. Thus a new methodology of geo-spatial analysis and the related theoretical models emerge. Based on the least squares regression and matrix operations, a simple algorithm is proposed to solve the multiscaling allometric equation. Applying the analytical method of multielement allometry to Chinese cities and regions yields satisfying results. A conclusion is reached that the multiscaling allometric analysis can be employed to make a comprehensive evaluation for the relative levels of urban and regional development, and explain spatial heterogeneity. The notion of multiscaling allometry may enrich the current theory and methodology of spatial analyses of urban and regional evolution.

## 1 Introduction

Allometric phenomena are ubiquitous in both nature and society, and we can find allometric scaling relations everywhere. In fact, the concept of allometry originated from biology, concerning the study of the relationship between size and shape (Gould, 1966; Lee, 1989). If the ratio of the relative rate of growth of an organ to that of another organ is a constant, we will say that there exists an allometric growth (Beckmann, 1958; Bertallanfy, 1968; Damuth, 2001; Small, 1996). The law of allometric growth was initially introduced into social science to research urbanization by Naroll and Bertalanffy (1956). Subsequently, the allometry idea was adopted to model the relationship between a system of cities and its largest city within a geographical region (Beckmann, 1958; Carroll, 1982; Chen, 2008a; Pumain and Moriconi-Ebrard, 1997; Pumain *et al*, 2006; Zhou, 1995). Since then, allometry has been attracting attention of urban geographers and city planners for many years, and a mass of studies on urban allometry were made (Arcaute *et al*, 2015; Batty and Longley, 1994; Chen, 2010; Lee, 1989; Lo and Welch, 1977; Longley, 1991; Nordbeck, 1971; Tobler, 1969). Among various studies on urban allometry, the works on the scaling relations between urban area and population size came into focus (Batty and Longley, 1994; Chen and Xu, 1999; Lo, 2002). The experimental results put the allometric analyses of cities in a dilemma of dimension because the empirically calculated values of the allometric power exponents always differ significantly from the theoretically expected values (Chen, 2008a; Lee, 1989). The ideas from fractal dimension raised the allometric models phoenix-like from the ashes (Chen, 2014a). Today, the allometric scaling is often associated with fractals (Batty and Longley, 1994; Chen, 2010; Enguist *et al*, 1998; He, 2006; West, 2002; West *et al*, 1997; West *et al*, 1999), and it has become one of basic laws in urban geography (Chen, 2014a; Lo, 2002). Allometric analysis can be applied to many fields of urban and regional researches (Batty *et al*, 2008; Bettencourt, 2013; Bettencourt *et al*, 2007; Bettencourt *et al*, 2010; Chen and Jiang, 2009; Kühnert *et al*, 2006; Lobo *et al*, 2013; Louf and Barthelemy, 2014a; Ortman *et al*, 2014; Samaniego and Moses, 2008; Zhang and Yu, 2010). Despite all those academic achievements, there is an important problem remaining to be solved, that is, it is necessary to find a way of integrating allometric scaling processes of multiple elements into a new theoretical framework.

Geographical research is involved with complexity science, and both cities and regions are complex spatial systems (Allen, 1997; Batty, 2008; Chen, 2008a; Portugali, 2011; Wilson, 2000). A geographical region is in fact an urban system comprising a network of cities and its hinterland. Generally speaking, an urban system can be divided into two levels. One is a system of cities (urban network) belonging to *interurban geography*, and the other is a city as a system (city system) belonging to *intraurban geography* (Batty and Longley, 1994; Berry, 1964). In a system of cities, each city develops; in a city as a system, each urban element grows. A problem is how to measure and compare the levels of development of different cities or various urban elements such as buildings, roads, and open space. Based on the concepts from fractal geometry (Mandelbrot, 1983) and general system theory (Bertallanffy, 1968), an *allometric scaling analysis* (ASA) method for evaluating city development was proposed by Chen (2008) and Chen and Jiang (2009). As far as the mathematical principle is concerned, this method bears an analogy with the *analytical hierarchy process* (AHP) developed by Satty (1999, 2008). The essential difference between ASA and AHP is as below: the former depends on the power exponent matrix coming from objective allometric analyses, while the latter relies on pairwise comparison matrix resulting from subjective judgment of decision-makers.

Though a preliminary approach and its basic mathematical principle have been advanced, several problems such as algorithms, statistical test of results, and application to geographical analysis need to be tackled. In this paper, the ASA method of cities will be further developed to form a method of *multiscaling allometry* (MSA). The analytical process of MSA will be illuminated, two test approaches will be put forward for MSA modeling, and a concise example will be illustrated so that readers will understand and be able to utilize this analytical process. The rest of this paper is arranged as below. In Section 2, the mathematical models of MSA will be presented, and the approach to estimating the parameters, the test methods for evaluating the modeling results, and an oversimplified example will be clarified. In Section 3, the MSA analysis will be applied to the 27 regions and 4 municipalities of China to make a typical case study. In Section 4, several related questions on the MSA method will be discussed. Finally, the paper will be concluded by outlining the major points of this work.

## 2 Models

### 2.1 A framework of MSA modeling

First of all, the precondition of the MSA analysis should be clarified. Allometry of cities and regions at least falls into two types: longitudinal allometry and transversal allometry (Pumain and Moriconi-Ebrard, 1997). The former is the allometric growth which can be investigated with time series, while the latter is the allometric distributions which can be examined through cross-section data and termed "cross-sectional allometry" (Chen, 2010; Chen, 2014a). The MSA analysis is based on longitudinal allometry, but it involves transversal relationships. Its object is continuous panel data, or multiple parallel time series. Thus, two postulates should be stated as follows. First, for a given measure, $x$, the allometric growth law dominates any pair of elements in a geographical system. Although the allometric scaling relations in the real world may partly degenerate from power laws, it can be approximately treated as allometric relation (Chen, 1995; Chen and Wang, 1997). Second, any element in the system undergoes non-negative growth, i.e., $dx/dt \geq 0$, where $t$ denotes time (Chen, 2008a; Chen and Jiang, 2009). Thus the scaling exponents are positive for ever. The allometric relations between geographical elements will be acceptable if a growth curve can be fitted to a pair of time series (Chen, 2014a).

In the spatio-temporal analysis of urban and regional systems, a comparison is often drawn between a part and the whole. This reminds us of the law of allometric growth in biology, which reads that the rate of relative growth of an organ (part) is a constant fraction of the rate of relative growth of the organism (whole) (Bertalanffy and Pirozynsky, 1952; Lee, 1989). By analogy with the biological allometry, Beckmann (1958) proposed an allometric model of urban systems, which asserts that the rate of relative growth of the central/largest city (a part) is a constant fraction of the rate of relative growth of the systems of cities (the whole) (Carroll, 1982; Zhou, 1995). This allometric relation has been empirically confirmed by Chinese, English, French, and Indian datasets of cities (Chen, 2008a; Pumain and Moriconi-Ebrard, 1997). Beckmann's model can be generalized to describe the allometric relation between an urban system (the whole) and any city (a part) in this system (Chen, 2008a). Suppose there is an urban system with $n$ cities. The allometric scaling relation between a city and the system of cities including the city can be expressed as

$$Q_i = a_i S^{b_i} = a_i (\sum_{i=1}^{n} Q_i)^{b_i},$$  (1)

where $Q_i$ refers to some measure (e.g. urban population size) of the $i$th city in the system, $a$ refers to the proportionality coefficient, and $b$, to the *allometric scaling exponent* (ASE) (Chen, 2010), $S$ denotes the sum of measurements of the $n$ cities, that is

$$S = \sum_{i=1}^{n} Q_i \,, \tag{2}$$

in which $i=1,2,\ldots,n$. In technique, the allometric model can be further generalized to describe the scaling relation between a set of cities (the whole) and any element within the set (a part).

The scaling exponent $b$ has two aspects of meaning: one is temporal meaning, and the other, spatial meaning. The temporal meaning is associated with the relative rate of growth (RRG) of cities. Taking the derivative of equation (1) with respect to time $t$ yields

$$b_i = (\frac{\mathrm{d}Q_i}{Q_i \mathrm{d}t}) / (\frac{\mathrm{d}S}{S\mathrm{d}t}) = \frac{\mathrm{d}Q_i / Q_i}{\mathrm{d}S / S} \,. \tag{3}$$

This implies that the $b_i$ value is the ratio of the relative growth rate of the $i$th city to that of the system of cities. If $b_i > 1$, the relative growth of the $i$th city is faster than that of the urban system as a whole (positive allometry); If $b_i < 1$, the relative growth of the $i$th city is slower than that of the urban system (negative allometry); If $b_i = 1$, the relative growth of the $i$th city is the same as that of the urban system (isometry). The spatial meaning is involved with the fractal dimension of cities (Chen, 2014b). There are two basic approaches to understanding fractals and fractal dimension. One is the scale-measure relation (spatial measurement process), for example, the power law relation between side lengths and numbers of nonempty boxes, the inverse power law relation between radius and density, and so on; the other is the measure-measure relation (geometric measure relation), for instance, the power law relation between area and perimeter, the inverse power law relation between rank and size, and the like (Mandelbrot, 1983; Feder, 1988). An allometric scaling relation is in fact a geometric measure relation (Feder, 1988; Takayasu, 1990), thus we have

$$Q_i^{D_{(i)}} \propto S^{D_s} \,, \tag{4}$$

where the symbol "$\propto$" means "be proportional to", $D_{(i)}$ refers to the fractal dimension of the $i$th city with respect to the measure $Q_i$, and $D_s$ refers to the fractal dimension of the measure $S$, which is defined by equation (2). The parameter $D_s$ reflects the overall effect of fractal dimension values of the $n$ cities. This suggests that the scaling exponent can be employed to evaluate the level of development of a city (part) relative to its urban system (the whole).

The fractal dimension suggests the level of space filling of a city or a system of cities in the process of urban evolution. Comparing equation (4) with equations (1) and equation (3) shows

$$b_i = \frac{\mathrm{d}Q_i / Q_i}{\mathrm{d}S / S} = \frac{D_{(i)}}{D_s},$$ (5)

which relates the time meaning of the allometric exponent with its space meaning. This suggests that the allometric exponent is the ratio of fractal dimensions. The fractal dimension indicates the extent of space filling, while the allometric scaling exponent can reflect the ratio of the space filling extent of one city to that of another city. Equation (5) indicates that the scaling exponent can be adopted to evaluate the relative level of development of a city in the system of cities comprising it. Using the regression analysis based on double logarithmic relation, we can obtain a scaling exponent vector $\mathbf{B}=[b_1 \ b_2 \ \dots \ b_n]^{\mathrm{T}}$. Unitizing the vector yields a set of indexes that can reflect the relative development levels of the $n$ cities.

The element-system (part-whole) allometric scaling is very simple, but it cannot reveal the deep structure of urban systems. In order to reflect the rich spatio-temporal information of city development, the allometric relation between the urban system and its elements can be converted into the relation between any two cities in the system. A basic assumption is that the ratio of the relative rate of growth of one element to that of another element approaches constant. In practice, the condition can be relaxed, thereby the tests become necessary. Thus we have element-element (part-part) allometric scaling. The allometric relation between city $i$ and city $j$ is as follows (Chen, 2008a; Chen and Jiang, 2009)

$$Q_i = \beta_j Q_j^{\alpha_{ij}} = \beta_j Q_j^{D_{(i)}/D_{(j)}},$$ (6)

where $Q_i$ and $Q_j$ refer to size measures such as urban population of the two cities, $\beta_j$ to the proportionality constant, $\alpha_{ij}=\mathrm{d}\ln Q_i/\mathrm{d}\ln Q_j$ to a scaling exponent, and $D_{(i)}$ and $D_{(j)}$ to the fractal dimension of $Q_i$ and $Q_j$, respectively. The fractal dimension can be understood through the geometric measure relation (Chen, 2014b; Feder, 1988; Takayasu, 1990). It can be proved that (Chen, 2008a; Chen and Jiang, 2009; Chen and Lin, 2009)

$$\alpha_{ij} = \frac{D_{(i)}}{D_{(j)}} = \frac{\mathrm{d}Q_i / Q_i}{\mathrm{d}Q_j / Q_j} = \frac{r_i}{r_j},$$ (7)

in which $\alpha_{ij}$ refers to the scaling exponent of the allometric relation between cities $i$ and $j$, $r_{(i)}=$ $\mathrm{d}\ln Q_i/\mathrm{d}t$ and $r_{(j)}=\mathrm{d}\ln Q_j/\mathrm{d}t$ are the relative rates of growth (RRG) of $Q_i$ and $Q_j$, here "ln" represents the function of natural logarithm. The allometric analysis of cities can be carried out by pairwise correlation of cities by means of the log-log linear regression based on equation (6). In other words, we can pair the cities and make allometric analyses pair by pair. The allometric exponents form a positive reciprocal matrix in the form (Chen, 2008a)

$$\mathbf{M} = \left[\alpha_{ij}\right]_{n \times n} = \begin{bmatrix} \alpha_{11} & \alpha_{12} & \cdots & \alpha_{1n} \\ \alpha_{21} & \alpha_{22} & \cdots & \alpha_{2n} \\ \vdots & \vdots & \cdots & \vdots \\ \alpha_{n1} & \alpha_{n2} & \cdots & \alpha_{nn} \end{bmatrix} = \left[D_{(i)} / D_{(j)}\right]_{n \times n}, \tag{8}$$

where $\mathbf{M}$ represents both the *scaling exponent matrix* (SEM) and the *fractal parameter matrix* (FPM) of urban allometric growth, the properties of the SEM are as below: $\alpha_{ii} = \alpha_{jj} = 1$, $\alpha_{ij} = 1/\alpha_{ji} = \alpha_{ik}/\alpha_{jk}$, where $i$, $j$, $k = 1, 2, \cdots, n$. Based on the SEM, a *fractal dimension matrix* (FDM) equation can be constructed as follows (Chen, 2008a; Chen and Jiang, 2009)

$$\mathbf{MD} = \begin{bmatrix} D_{(1)}/D_{(1)} & D_{(1)}/D_{(2)} & \cdots & D_{(1)}/D_{(n)} \\ D_{(2)}/D_{(1)} & D_{(2)}/D_{(2)} & \cdots & D_{(2)}/D_{(n)} \\ \vdots & \vdots & \cdots & \vdots \\ D_{(n)}/D_{(1)} & D_{(n)}/D_{(2)} & \cdots & D_{(n)}/D_{(n)} \end{bmatrix} \begin{bmatrix} D_{(1)} \\ D_{(2)} \\ \vdots \\ D_{(n)} \end{bmatrix} = n \begin{bmatrix} D_{(1)} \\ D_{(2)} \\ \vdots \\ D_{(n)} \end{bmatrix} = n\mathbf{D}, \tag{9}$$

where $\mathbf{D} = [D_{(1)}\, D_{(2)} \ldots D_{(n)}]^{\mathrm{T}}$ denotes a *fractal dimension vector* (FDV). If a city bears multifractal structure rather than monofractal form, we only consider the capacity dimension based on the 0 order of moment. Apparently, $n$ is just the maximum eigenvalue of $\mathbf{M}$, and $\mathbf{D}$ is the corresponding eigenvectors (Chen, 2008a). Normalizing the eigenvector yields

$$d_i = D_{(i)} / \sqrt{\sum_{i=1}^{n} D_{(i)}^2} \,. \tag{10}$$

Thus an element-element *allometric scaling index* (ASI) can be defined as

$$w_i = d_i / \sum_{i=1}^{n} d_i = D_{(i)} / \sum_{i=1}^{n} D_{(i)} \,, \tag{11}$$

which is the unitized eigenvector. We can evaluate a city's relative development level by means of the ASI values. ASI provides a simple measurement that is concentrated on a concise number of the relative growth information of many parts in a system. The common allometric scaling exponent such as $b_{ij}$ in equation (3) is used to compare the relative rates of growth of two correlative parts. Differing from the scaling exponent $b_{ij}$, the ASI defined by equations (9) and (10) can reflect the growth rate of one part relative to all other parts in a system.

The element-element allometry can be mathematically associated with the element-system allometry of cities. A new finding is that there exists an approximate relation as follows

$$\hat{w}_i^* = b_{(i)} / \sum_{i=1}^{n} b_{(i)} \approx \hat{w}_i \,,$$ (12)

where $w_i^*$ is an approximate of $w_i$. It is an element-system scaling index and can be termed *approximate allometric scaling index* (AASI). For geographers, it is hard to grasp the meaning of ASI, but it is easy to understand the scaling exponent, $b_i$. Equation (12) indicates a simple and clear way of understanding the ASI. According to the approximate relationship, the part-part allometric scaling and the part-whole allometric scaling can be converted into each other. Thus, the part-part allometric relations based on equation (6) and the part-whole allometric relations based on equation (1) combines to make a new methodology termed *MSA analysis* for development evaluation of geographical systems. The analytical process can be illustrated as follows (Figure 1).

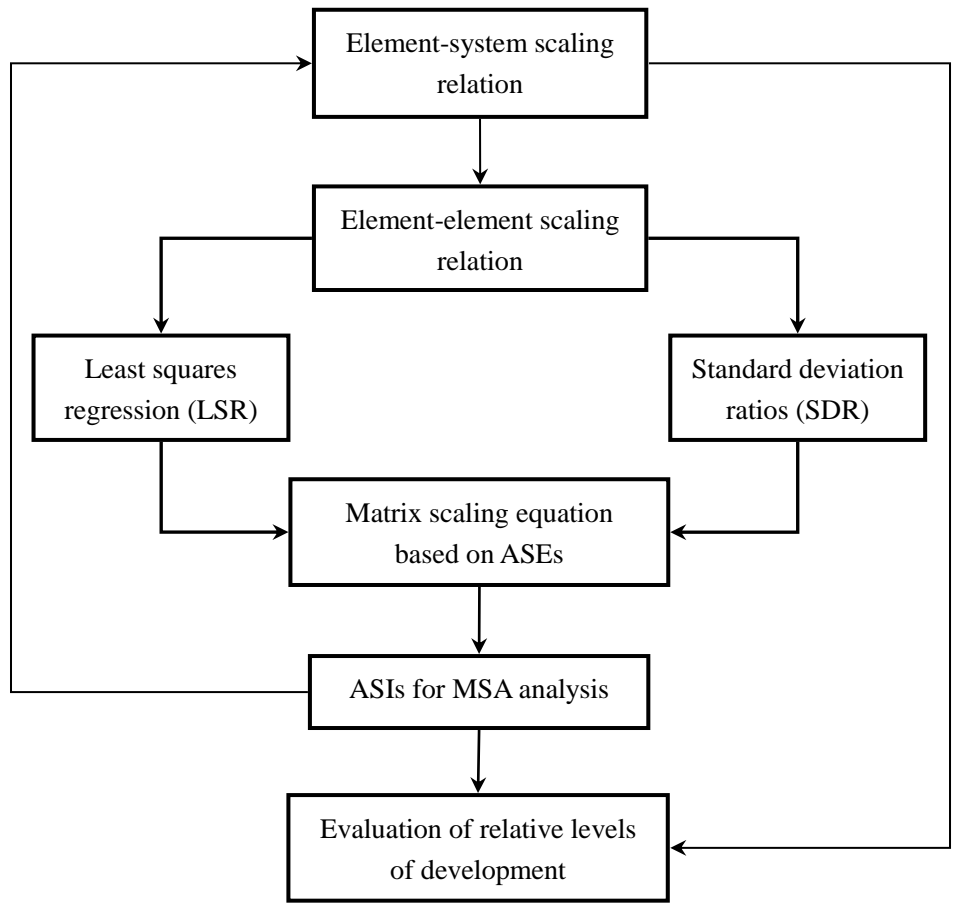

**Figure 1 A schematic diagram of the MSA analysis of urban and regional development**

The above models are theoretical expressions rather than practical approaches. The allometric scaling relations of cities, equations (1) and (6), are based on the following assumption: each pair of cities in an urban system follows the law of allometric growth. If the urban growth complies with the allometric scaling law absolutely, it can be proved that the entries of the SEM would meet the following reciprocal condition

$$\alpha_{ij} = \frac{D_{(i)}}{D_{(j)}} = \frac{r_i}{r_j} = \frac{\sigma_i}{\sigma_j} = \frac{1}{\alpha_{ji}}, \tag{13}$$

in which $\sigma_i$ and $\sigma_j$ denote the theoretical standard deviations of $\ln Q_i$ and $\ln Q_j$ (see demonstration in MMC1 File). Equation (13) relates the spatial meaning ($D$) and temporal meaning ($r$) of the scaling exponent to its information meaning ($\sigma$). The relative growth rate ($r$) is a measure of space-filling process, while the self-similar fractal dimension ($D$) is a measure of space-filling pattern. A geographical pattern is always associated with the corresponding geographical process. The concept of allometric growth suggests different relative growth rates, which further suggests different space-filling extents. The higher relative growth rate leads to the higher space-filling extent, and thus results in a higher fractal dimension; the lower relative growth rate leads to the lower space-filling extent, and thus results in a lower fractal dimension.

If a spatial pattern is fractal, the corresponding temporal process is also fractal, and the fractal dimension hidden in the time series can be estimated by the methods of phase space reconstruction and multidimensional scaling analysis. The procedure is as follows. First, by reconstructing phase space based on time-lag effect, we can obtain a distance matrix, from which we can calculate the correlation dimension (Chen, 2012a; Kantz and Schreiber, 1997; Packard *et al*, 1980; Takens, 1981; Williams, 1997). Then, by means of Tobler's multilateration method for multidimensional scaling (Golledge and Rushton, 1972; Haggett, 2001; Haggett *et al*, 1977), we can convert the distance matrix based on state space into a map defined in a 2-dimension real space (Chen, 2008a). Finally, some methods such as box counting can be employed to estimate the capacity dimension through the map proceeding from the fractal time series. What is more, power spectral analysis can be employed to calculate the Hurst exponent, from which we can derive the self-similar dimension of a nonlinear time series of urban evolution (Chen, 2013).

However, in most cases, it is either unnecessary or impossible to compute the fractal dimension using a time series for spatial analysis. On the one hand, the significance and value of a measure or dimension rest with comparison. Compared with another fractal dimension value, a fractal dimension value can be better brought to light and reflect the degree of space filling efficiently. On the other, if the sample path of a time series is short, we cannot figure out the fractal dimension of a geographical process. The allometric scaling analysis can help us to overcome the difficulties abovementioned. It is easier to calculate an allometric scaling exponent, which, as a fractal dimension ratio, takes on comparative meaning of different fractal parameters.

In empirical studies, an allometric exponent is not equal to its reciprocal value because of random disturbance of observation and computation or the random deviation from allometric scaling relation. In particular, due to space-time translational asymmetry of geographical mathematical laws (Chen, 2008a; Chen, 2014b), we cannot guarantee that any pair of cities in a network of cities always follows the allometric growth law. By the principle of least squares method, we can derive the following relations

$$\hat{a}_{ij} = R_{ij} \frac{s_i}{s_j}, \hat{a}_{ji} = R_{ji} \frac{s_j}{s_i},$$  (14)

where $R_{ij}$ denotes the correlation coefficient of the allometric relation between cities $i$ and $j$, $s_i$ and $s_j$ are the sample standard deviations of $\ln Q_i$ and $\ln Q_j$, and the hat symbol '^' implies estimation. If $R_{ij} = 1$, then equations (14) will return to equations (13). Because of the symmetry of correlation, i.e., $R_{ij} = R_{ji}$, from equations (14) it follows

$$\hat{\alpha}_{ij} = \frac{R_{ij}^2}{\hat{\alpha}_{ji}},$$  (15)

where $R^2$ denotes the goodness of fit (GOF) of the linear regression modeling (demonstration in MMC1 File). This suggests that equation (13) should be substituted by equation (15) in practice. Thus, according to equations (13) and (14), the theoretical relation expressed by equation (9) should be replaced by

$$\hat{\mathbf{M}}\hat{\mathbf{D}} = \begin{bmatrix} R_{11}s_1/s_1 & R_{12}s_1/s_2 & \cdots & R_{1n}s_1/s_n \\ R_{21}s_2/s_1 & R_{22}s_2/s_2 & \cdots & R_{2n}s_2/s_n \\ \vdots & \vdots & \cdots & \vdots \\ R_{n1}s_n/s_1 & R_{n2}s_n/s_2 & \cdots & R_{nn}s_n/s_n \end{bmatrix} \begin{bmatrix} \hat{D}_1 \\ \hat{D}_2 \\ \vdots \\ \hat{D}_n \end{bmatrix} = \lambda_{\max} \hat{\mathbf{D}},$$  (16)

where $\mathbf{M}$-hat and $\mathbf{D}$-hat represent the empirical results of $\mathbf{M}$ and $\mathbf{D}$, which differ from the theoretical values of the fractal dimension matrix and vector, $\lambda_{\max}$ refers to the largest eigenvalue corresponding to **D**-hat. The maximum eigenvalue $\lambda_{max}$ is used to approximate the number of elements, $n$. In theory, $\lambda_{max} = n$, but in empirical analyses, $\lambda_{max} \to n$. The closer $R_{ij}^2$ is to 1, the closer $\lambda_{max}$ is to $n$, and in turn the closer **D**-hat is to **D**. Based on equation (16), a new simple algorithm can be developed for the MSA model.

## 2.2 Algorithms and a simple example

The precondition of application of a model to actual problems is to find effective algorithms. The keys to making a MSA analysis rest with two procedures: one is to evaluate the SEM, equation (8), and the other, is to solve the FDM equation, equation (9). As indicated above, equation (9) should be substituted by equation (16) in practice. The least squares regression (LSR) can be employed to evaluate the SEM. However, it is inconvenient to use the LSR method to obtain the SEM in practice. In this paper, a simple and thus accessible approach is presented. This approach is based on the matrix and array multiplication of the correlation coefficients and standard deviations. The main is the ratios of the standard deviations of the logarithmic variables, so the algorithm can be briefly termed *standard deviation ratios* (SDR) method.

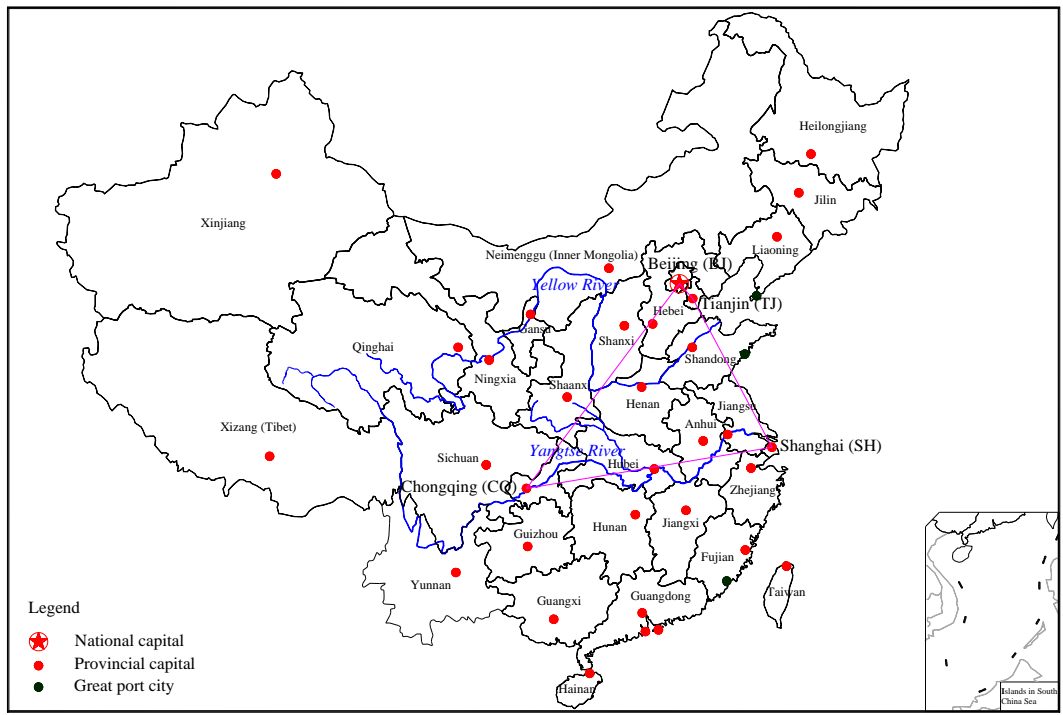

**Figure 2 A sketch map of Chinese cities and regions including the provinces, autonomous regions, and municipalities directly under the Central Government**

For simplicity, a small set of cities can be taken for an example to illustrate the process of calculation. One of functions of the MSA analysis is to evaluate the relative levels of city development and generate a rank of the growth potential for cities. A MSA analysis can be made for the four municipalities directly under the Central Government of China, Beijing (BJ), Tianjin (TJ), Shanghai (SH), and Chongqing (CQ) (Figure 2). The basic size measurement is *gross regional product* (GRP), the period is from 1998 to 2012, and thus the length of sample path is $T$=15 (Figure 3, Table 1). Now, the approach of SDR comprising four steps is used to determine the SEM of the four cities (datasets in MMC2 File).

The first step is to take the logarithm of the observational data by time and urban elements. This step is simple. The formula is

$$x_{tj} = \ln(Q_{tj}),\qquad(17)$$

where $t$=1,2,…,$T$ refers to time, and $T$ to the length of sample path ($j$=1,2,…, $n$; $T$=15). The results form a vector $\mathbf{x}_j$=[$x_{1j}$ $x_{2j}$ … $x_{Tj}$]$^{\mathrm{T}}$, which in turn make a matrix of logarithmic variables such as $\mathbf{X}$=[$x_1 x_2 … x_n$].

The second step is to standardize the logarithmic variables by urban elements. This step is still simple. The formula is

$$\mathbf{y}_j = \frac{\mathbf{x}_j - \bar{x}_j}{s_j} = \frac{\ln(Q_j) - \bar{x}_j}{s_j},\qquad(18)$$

where $\bar{x}_j$ denotes the average value of $x_j$, and $s_j$ is the corresponding standard deviation. The results form a standardized matrix of logarithmic variables $\mathbf{Y}$=[$y_1 y_2 … y_n$] (Table 1).

The third step is to compute the correlation coefficients based on the logarithmic linear relations. It is easy to reckon the Pearson correlation coefficients using matrix multiplication. On the basis of *sample standard deviation* (SSD), the formula is

$$\mathbf{V} = \frac{1}{n-1}\mathbf{Y}^{\mathrm{T}}\mathbf{Y} = \left[R_{ij}\right]_{n\times n} = \begin{bmatrix} R_{11} & R_{12} & \cdots & R_{1n} \\ R_{21} & R_{22} & \cdots & R_{2n} \\ \vdots & \vdots & \cdots & \vdots \\ R_{n1} & R_{n2} & \cdots & R_{nn} \end{bmatrix},\qquad(19)$$

where $R_{ij}$ is the coefficient of Pearson correlation between city $i$ and city $j$ ($R_{ii}$=$R_{jj}$=1). If we use the *population standard deviation* (PSD) to standardize the random variables for theoretical analyses, equation (19) should be replaced by $\mathbf{V}$=$\mathbf{Y}^{\mathrm{T}}\mathbf{Y}$/$n$ (Table 2).

The fourth step is to evaluate the ASEs using correlation coefficients and logarithmic standard deviations. The formula is equations (14), and the result is

$$\hat{\mathbf{M}} = \left[\hat{\alpha}_{ij}\right]_{n \times n} = \begin{bmatrix} R_{11}s_1/s_1 & R_{12}s_1/s_2 & \cdots & R_{1n}s_1/s_n \\ R_{21}s_2/s_1 & R_{22}s_2/s_2 & \cdots & R_{2n}s_2/s_n \\ \vdots & \vdots & \cdots & \vdots \\ R_{n1}s_n/s_1 & R_{n2}s_n/s_2 & \cdots & R_{nn}s_n/s_n \end{bmatrix},$$ (20)

which is quasi-reciprocal matrix rather than a real reciprocal matrix (Table 2).

**Table 1 The transformed results of the gross regional product (GRP) of four Chinese municipalities (1998-2012)**

| Year | Logarithmic variable | | | | Standardized variable | | | |
|------|---------|---------|----------|------------|------------|------------|------------|------------|
| | Beijing | Tianjin | Shanghai | Chongqing | Beijing | Tianjin | Shanghai | Chongqing |
| | $\ln(Q_1)$ | $\ln(Q_2)$ | $\ln(Q_3)$ | $\ln(Q_4)$ | $\ln(Q_1)^*$ | $\ln(Q_2)^*$ | $\ln(Q_3)^*$ | $\ln(Q_4)^*$ |
| 1998 | 7.9434 | 7.2885 | 8.2926 | 7.3967 | -1.4855 | -1.3779 | -1.4914 | -1.2629 |
| 1999 | 8.0214 | 7.3702 | 8.3825 | 7.4314 | -1.3594 | -1.2657 | -1.3296 | -1.2100 |
| 2000 | 8.1523 | 7.4929 | 8.5028 | 7.5029 | -1.1477 | -1.0971 | -1.1128 | -1.1010 |
| 2001 | 8.2904 | 7.6084 | 8.5870 | 7.5991 | -0.9246 | -0.9383 | -0.9613 | -0.9543 |
| 2002 | 8.4117 | 7.7170 | 8.6755 | 7.7183 | -0.7286 | -0.7891 | -0.8020 | -0.7725 |
| 2003 | 8.5602 | 7.8982 | 8.8291 | 7.8512 | -0.4885 | -0.5401 | -0.5254 | -0.5699 |
| 2004 | 8.7478 | 8.0861 | 9.0163 | 8.0207 | -0.1853 | -0.2818 | -0.1882 | -0.3113 |
| 2005 | 8.8493 | 8.2702 | 9.1321 | 8.1513 | -0.0213 | -0.0288 | 0.0202 | -0.1122 |
| 2006 | 9.0018 | 8.4035 | 9.2660 | 8.2706 | 0.2252 | 0.1544 | 0.2613 | 0.0698 |
| 2007 | 9.1949 | 8.5665 | 9.4330 | 8.4502 | 0.5373 | 0.3785 | 0.5620 | 0.3437 |
| 2008 | 9.3161 | 8.8127 | 9.5518 | 8.6645 | 0.7331 | 0.7168 | 0.7759 | 0.6705 |
| 2009 | 9.4053 | 8.9256 | 9.6189 | 8.7842 | 0.8774 | 0.8719 | 0.8967 | 0.8530 |
| 2010 | 9.5549 | 9.1296 | 9.7507 | 8.9779 | 1.1192 | 1.1524 | 1.1340 | 1.1484 |
| 2011 | 9.6960 | 9.3332 | 9.8624 | 9.2115 | 1.3472 | 1.4322 | 1.3352 | 1.5047 |
| 2012 | 9.7914 | 9.4645 | 9.9125 | 9.3422 | 1.5014 | 1.6126 | 1.4254 | 1.7041 |
| Mean | **8.8625** | **8.2911** | **9.1209** | **8.2248** | **0.0000** | **0.0000** | **0.0000** | **0.0000** |
| SSD (*s*) | **0.6187** | **0.7276** | **0.5554** | **0.6557** | **1.0000** | **1.0000** | **1.0000** | **1.0000** |

**Note**: The original data come from National Bureau of Statistics of China (http://www.stats.gov.cn/tjsj/ndsj/). Unit: 100 millions *yuan* (RMB). The abbreviation SSD denotes "sample standard deviation".

**Table 2 The matrices of the correlation coefficients and allometric power exponents**

| City | Correlation coefficient matrix | | | | Allometric exponent matrix | | | |
|------|--------|--------|--------|--------|--------|--------|--------|--------|
| | BJ | TJ | SH | CQ | BJ | TJ | SH | CQ |
| BJ | 1 | 0.9968 | 0.9992 | 0.9914 | 1 | 0.8476 | 1.1131 | 0.9355 |
| TJ | 0.9968 | 1 | 0.9956 | 0.9982 | 1.1722 | 1 | 1.3043 | 1.1077 |
| SH | 0.9992 | 0.9956 | 1 | 0.9891 | 0.8969 | 0.7599 | 1 | 0.8377 |
| CQ | 0.9914 | 0.9982 | 0.9891 | 1 | 1.0507 | 0.8996 | 1.1677 | 1 |

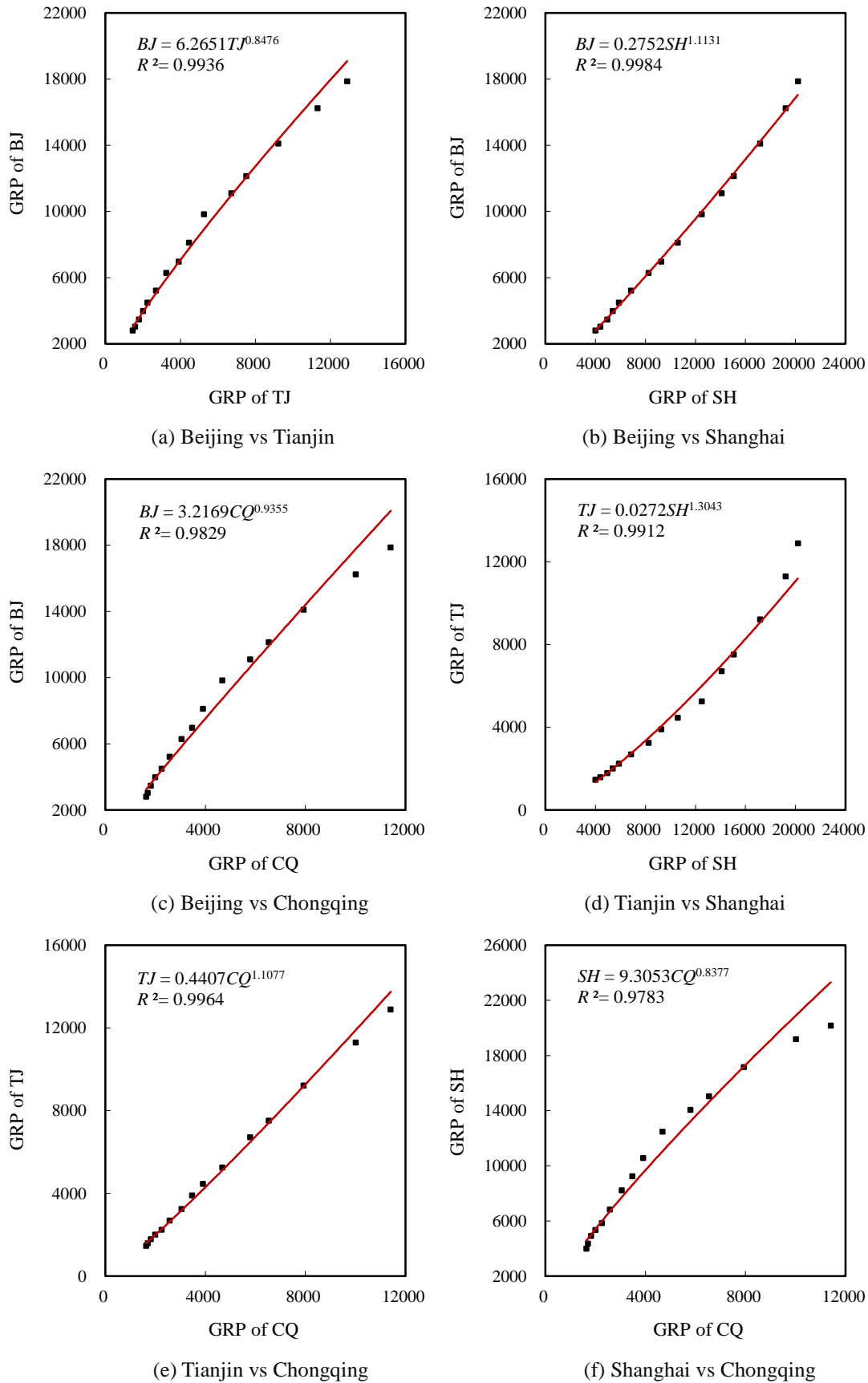

**Figure 3 The allometric scaling relations between cities of China in terms of GRP (1998-2012)**

**Note**: The numerical unit of the coordinate axes is "100 millions *yuan* (RMB)".

The second procedure is to work out the FDM equation, and this aim can be accomplished by finding the eigenvector of the SEM corresponding to the maximum eigenvalue $n$. At least three approaches can be used to calculate the largest eigenvalue and the corresponding eigenvector. The first is the matrix power method, the second is the arithmetical averaging method, and the third is the geometric averaging method. Among the three methods, the geometric averaging is the best way of approximate estimation: the process is simple and the result is credible (Chen, 2008a). Suppose that the SEM has been obtained by the method of LSR or SDR (Table 2). For the four cities abovementioned, the geometric averaging approach and results are as follows. The first step is to calculate the geometric average values by row. The formula and results are as below

$$\hat{\mathbf{W}} = (\prod_{j=1}^{n} \hat{\alpha}_{ij})^{1/n} = \begin{bmatrix} 0.9693 & 1.1408 & 0.8693 & 1.0250 \end{bmatrix}^{\mathrm{T}}, \tag{21}$$

which is an approximation of the eigenvector of the SEM. The second step is to compute the eigenvector by normalizing the geometric average values. The formula and results are as follows

$$\hat{\mathbf{U}} = \hat{W}_i / (\sum_{i=1}^{n} W_i^2)^{1/2} = \begin{bmatrix} 0.4818 & 0.5671 & 0.4321 & 0.5095 \end{bmatrix}^{\mathrm{T}}, \tag{22}$$

which can be treated as a normalized eigenvector of the SEM. The third step is to unitize the normalized eigenvector. The formula and results are as below

$$\hat{\mathbf{w}} = U_i / \sum_{i=1}^{n} U_i = \begin{bmatrix} 0.2421 & 0.2849 & 0.2171 & 0.2560 \end{bmatrix}^{\mathrm{T}}, \tag{23}$$

which indicates the relative shares of urban spatial competition or the relative levels of city development (Table 3). The fourth step is to reckon the maximum eigenvalue. The formula and result are

$$\lambda_{\max} = \sum_{i=1}^{n} (\hat{\mathbf{M}}\hat{\mathbf{w}})_i = 3.9851, \tag{24}$$

which approximates to the dimension of the corresponding eigenvector. In terms of equation (11), the second step can be skipped over in an empirical analysis. This step is preserved owing to theoretical consideration.

**Table 3 The matrices of allometric scaling exponents, geometric means, and unitized eigenvector**

| City | BJ | TJ | SH | CQ | Geometric mean ($W_i$) | ASI ($\hat{w}_i$) |
|------|------|------|------|------|------|------|
| BJ | 1 | 0.8476 | 1.1131 | 0.9355 | 0.9693 | 0.2421 |
| TJ | 1.1722 | 1 | 1.3043 | 1.1077 | 1.1408 | 0.2849 |
| SH | 0.8969 | 0.7599 | 1 | 0.8377 | 0.8693 | 0.2171 |
| CQ | 1.0507 | 0.8996 | 1.1677 | 1 | 1.0250 | 0.2560 |

It can be demonstrated that the element-element allometric scaling exponent (ASI) is proportional to element-system allometric scaling exponent (AASI). For the four cities, the part-whole scaling exponent, $b_i$, can be estimated by regression analysis. Unitizing the $b_i$ values, we can calculate the AASI, $\hat{w}_i^*$. Obviously, the AASI values $\hat{w}_i^*$ and the ASI values $\hat{w}_i$ are close in number to one another (Table 4). It is easy to program the computer to fulfill this series of calculations instead of manual operation (see programs in MMC3 File and MMC4 File).

Table 4 Comparison between the element-element allometric scaling exponents (ASIs) and the element-system allometric scaling exponents (AASIs) (1998-2012)

| City | Scaling exponent ($b_i$) | GOF ($R^2$) | AASI ($\hat{w}_i^*$) | ASI ($\hat{w}_i$) |
|------|------|------|------|------|
| Beijing (BJ) | 0.9989 | 0.9985 | 0.2422 | 0.2421 |
| Tianjin (TJ) | 1.1745 | 0.9981 | 0.2848 | 0.2849 |
| Shanghai (SH) | 0.8961 | 0.9971 | 0.2173 | 0.2171 |
| Chongqing (CQ) | 1.0546 | 0.9907 | 0.2557 | 0.2560 |

## 2.3 Tests and evaluation

The MSA provides a practical approach to modeling spatio-temporal evolution of urban and regional systems. Generally speaking, a model can be defined as proper "simplification of reality" (Longley, 1999). All mathematical modeling have two major functions: one is explanation, and the other, prediction (Kac, 1969; Fotheringham and O'Kelly, 1989). Any model has its valid scope of application. Beyond the scope, a model cannot perform its function. In order to judge whether a model can explain and predict reality, we must make necessary tests for the modeling result before applying it to actual problems. A test is in fact an evaluation of modeling quality from a given angle of view. By means of statistic tests, we can give a confidence statement about a conclusion on the basis of certain significance level. Facing a growing process, we don't know whether it follows the law of allometry. In this case, we should make tests for the results of an allometric analysis. Based on equation (16), two test methods can be developed for the MSA modeling.

The first test is based on the maximum eigenvalues of SEM. If and only if the allometric growth law is absolute, equation (13) will come into being and result in an equation $\lambda_{max} = n$, where $\lambda_{max}$ denotes the non-negative largest eigenvalue of SEM. That is to say, under the ideal condition, the element number $n$ is just the largest eigenvalue $\lambda_{max}$, and we have

$$\lambda_{max} = \sum_{i=1}^{n} \lambda_i = \sum_{i=1}^{n} \alpha_{ii} = n, \qquad (25)$$

where $\lambda_i$ represents the $i$th eigenvalue. Equation (25) lays the foundation for the first statistical test of the MSA analysis. Defining a *scaling consistency index* (SCI) as below

$$SCI = \frac{|\lambda_{max} - n|}{n-1}, \qquad (26)$$

we have $SCI$=(3.9851-4)/3=0.0050 for the four Chinese cities. The SEM bears an analogy with the pairwise comparison matrix of AHP propounded by Saaty (1999, 2008). So the test for positive reciprocal matrix consistency can be adopted for reference. By the results of random experiments, for $n$=4, the *random consistency index* (RCI) is about $RCI$=0.904 (Table 5). Thus, the *scaling consistency ratio* (SCR) is $SCR$=$SCI$/$RCI$=0.0055<<0.1. The value is small and the SEM empirically passed the scaling consistency test.

**Table 5 The values of the RCI for the scaling consistency test of the MSA analysis**

| $n$ | RC | $n$ | RC | $n$ | RC | $n$ | RC | $n$ | RC |
|---|---|---|---|---|---|---|---|---|---|
| 1 | 0 | 11 | 1.517 | 21 | 1.655 | 31 | 1.700 | 41 | 1.723 |
| 2 | 0 | 12 | 1.542 | 22 | 1.661 | 32 | 1.703 | 42 | 1.725 |
| 3 | 0.514 | 13 | 1.563 | 23 | 1.667 | 33 | 1.706 | 43 | 1.726 |
| 4 | 0.904 | 14 | 1.581 | 24 | 1.673 | 34 | 1.709 | 44 | 1.728 |
| 5 | 1.115 | 15 | 1.596 | 25 | 1.678 | 35 | 1.711 | 45 | 1.729 |
| 6 | 1.246 | 16 | 1.609 | 26 | 1.682 | 36 | 1.713 | 46 | 1.731 |
| 7 | 1.336 | 17 | 1.620 | 27 | 1.686 | 37 | 1.716 | 47 | 1.732 |
| 8 | 1.400 | 18 | 1.630 | 28 | 1.690 | 38 | 1.718 | 48 | 1.733 |
| 9 | 1.449 | 19 | 1.639 | 29 | 1.694 | 39 | 1.720 | 49 | 1.735 |
| 10 | 1.487 | 20 | 1.647 | 30 | 1.697 | 40 | 1.721 | 50 | 1.736 |

**Note**: The SCI values depend on sample sizes. The bigger a sample is, the stronger the random disturbance will be; and the stronger the random distribution is, the larger the SCI value will be. The random consistency indexes are obtained by random experiments to calibrate a SCI value so that we can make a credible judgment.

The second test is to make use of the correlation coefficient matrix. Based on equations (15) and (16), an *average correlation coefficient* (ACC) can be defined as

$$R^* = \frac{1}{(n-1)n}(\sum_{i=1}^{n}\sum_{j=1}^{n}R_{ij} - n)\,,\tag{27}$$

which is based on the symmetry of the correlation coefficient matrixes. Applying equation (27) to the correlation coefficients in Table 2 (left part) yields an ACC value $R^*$=0.9950 for the four Chinese municipalities discussed above. Then, the correlation coefficient test of regression analysis can be employed to evaluate the allometric scaling modeling. This test depends on the level of significance and degree of freedom. Input the formula "=(FINV($\alpha$,1,df)/(df+FINV($\alpha$,1, df)))^0.5" into any cell in a sheet of MS Excel, we can gain the critical value of $R^*$. Here "df" denotes the degree of freedom. For our example, the degree of freedom is df=$T$-2=13. If the significance level is taken as $a$=0.01, we can find a critical value $R_{c}^*$ =0.6411 by applying the formula "=(FINV(0.01,1,13)/(13+FINV(0.01,1,13)))^0.5" to Excel. Since $R^*$= 0.9950>0.6411, the ACC can pass the test of the confidence level of 99%.

The allometric scaling index is used to characterize the relative level of urban or regional development. In fact, we can employ the characteristic values of GRP within the 15 year to describe the absolute levels of economic development of the four cities. In the simplest case, the allometric growth can be derived from a pair of processes of exponential growth (Bertalanffy, 1968). Corresponding to equation (6), the exponential growth model can be expressed as

$$Q(t) = Q_0 \exp(\frac{t}{t_0})\,,\tag{28}$$

where $t_0$ denotes characteristic time length, and $Q_0$ refers to the initial value of the size measurement $Q$ ($t$=0). If $t$= $t_0$, we will have a characteristic size $Q_c$= $Q_0e$, where $e$≈2.7183. Using equation (28), we can calculate the characteristic values of GRP indicative of absolute development levels of the four cities (Table 6). Where relative level is concerned, Tianjin is the best one. Beijing is higher than Chongqing, and Shanghai is at the floor level. However, where absolute development level is concerned, Shanghai is higher than Beijing, which is in turn higher than Chongqing, and Tianjin is at the end of the rank (Figure 4). Comparatively speaking, among the four cities, Tianjin has a larger space to develop in future.

**Table 6 The initial observed values, initial predicted values, mean values, characteristic values, and goodness of fit of exponential models of GRP of four Chinese cities (1998-2012)**

| City | Observed value | | Predicted value | | Goodness of fit |
|------|---------------|------|----------------|------|-----------------|
| | $Q_{1998}$ | Mean | $Q_0$ | $Q_c$ | $R^2$ |
| Beijing (BJ) | 2816.8182 | 8385.4571 | 2684.1657 | 7296.3188 | 0.9977 |
| Tianjin (TJ) | 1463.4446 | 5089.1254 | 1281.8494 | 3484.4279 | 0.9933 |
| Shanghai (SH) | 3994.1895 | 10503.4641 | 3843.2280 | 10446.9768 | 0.9943 |
| Chongqing (CQ) | 1630.6707 | 4580.7380 | 1349.2924 | 3667.7571 | 0.9828 |

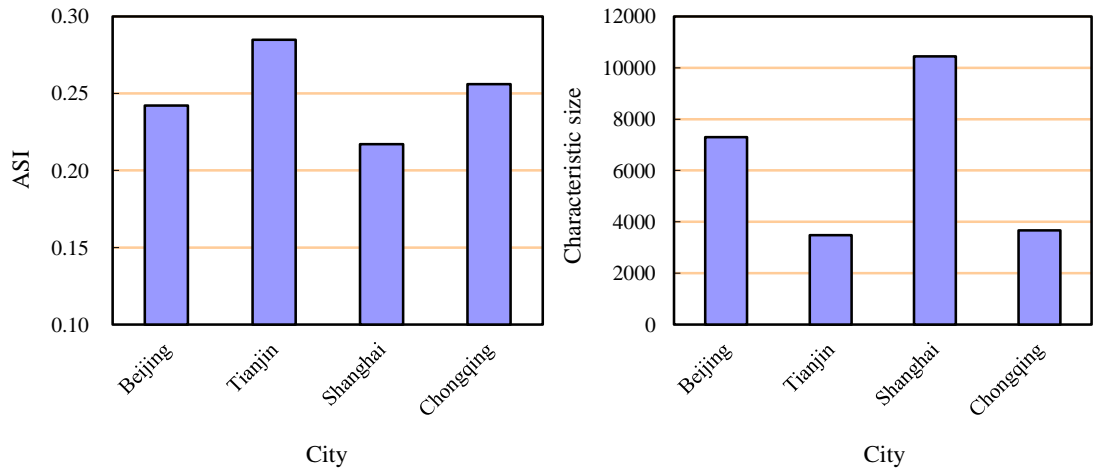

(a) Relative development level          (b) Absolute development level

**Figure 4 The relative development levels and absolute sizes of the four Chinese cities measured by average GRP and ASI (1998-2012)**

The basic principle of MSA rests with the allometric scaling and fractal property of urban systems. Both cities and regions are self-organizing systems with spatial complexity (Allen, 1997; Batty, 2005; Chen, 2008a; Portugali, 2000; Portugali, 2011; Wilson, 2000). Allometry and fractality occur often in complex spatial systems such as cities and networks of cities. Based on the scaling concept, allometric growth theory may be integrated with fractal geometry and complex network science to form a new theory about how cities and regions evolve from the bottom up (Batty, 2008; Chen, 2008a). The MSA analysis comes from interurban network analysis. This methodology can be developed at two aspects: one is to expand its scope of application, and the other, to improve the method itself. First, it can be applied to intraurban structure and regional systems. For example, the method can be utilized to evaluate the relative potential of different industrial sections in a city. We can rank the development level of the primary industry, secondary industry, and tertiary industry of Beijing or Shanghai. The analytical results are helpful for the selection of leading industries and determination of urban growth pole. The method can also be used to analyze spatio-temporal evolution of different types of urban land use, including residential land, industrial land, transportation land, municipal utility land, green land, and open space. Moreover, the ideas from allometric scaling can be used to reveal the relationships between a fractal set and its complementary set of urban form. It is well-known that the complement set of a fractal suggests a Euclidean dimension (Mandelbrot, 1982). For a city, build-up land and non-build up land are complementary sets. If urban build-up land is treated as a fractal set, the non-build up land as a complementary set is not fractal. The allometric scaling can be applied to fractal complements of cities. Second, the MSA method can be generalized to regional system analysis. For example, we can use it to research the 32 Chinese regions, including provinces, autonomous regions, and the municipalities directly under the Central Government of China.

The MSA analysis is a mono-variable modeling approach based on one measure of cities for the time being. The method can be readily improved by taking into account multiple variables. If we adopt different measures (e.g., population, transport, GRP, level of urbanization) to carry out MSA analyses on cities as systems or a system of cities, we can develop a multivariable multilevel MSA methodology for urban and regional development (Chen, 2008a). Based on the cascade structure of urban systems (Chen, 2016), the longitudinal allometry of different years and the transversal allometry of different elements can be integrated into a comprehensive analytical framework. The longitudinal allometry faces urban growth, while the cross-sectional allometry faces network of elements. The difficulty is how to find the observational data with high quality.

# 3 Empirical analysis

## 3.1 Study area and measurements

The case presented above is based on a small dataset of cities. The simple example helps us understand the computational and analytical processes. However, maybe the set of spatial elements is too small to convince readers of the effect of the MSA analysis. In fact, it is easy to generalize the method of MSA to analyze a larger spatial system. Next, the multi-element allometric analysis will be applied to the 31 cities and regions of Mainland China (Figure 2). Two measurements are adopted in this case study. One is GRP, the period is from 1998 to 2012 ($T$=15); and the other is level of urbanization, the period is from 2005 to 2013 ($T$=9). The original data come from the National Bureau of Statistics of China (see the datasets in MMC2 File). GRP can be used to evaluate the regional economic growth, and urbanization level can be used to evaluate the socio-economic development of different regions.

## 3.2 Calculations and analysis

The allometric indexes of GRP growth can be calculated using the abovementioned algorithm. Based on the element-element allometry, the ASIs are computed to reflect the local-local scaling relations; based on the element-system allometry, the AASIs are computed to reflect the global -local scaling relations. Within 3 decimal places, the ASIs and the AASI are almost the same with each other. From the 4th digits after the decimal points, the numerical differences begin to emerge (Table 7). The maximum characteristic root corresponding to the vector of ASI is about $\lambda_{max}$=30.8983, which results in $SCI$=0.0034. Given $n$=31, it follows $RCI$=1.7 (Table 5). Thus the ratio of scaling consistency is $SCR$=0.002< 0.1. The scaling consistency reaches the empirical standard. For the significance level $\alpha$=0.01 and degree of freedom $df$=13, the threshold value of Pearson correlation is $R_c^*$=0.6411. The ACC is about $R^*$=0.9966, which is great than $R_c^*$. This suggests that the confidence level of the MSA analysis is higher than 99%. In the regional system, the ASI values of the four municipalities, Beijing, Tianjin, Shanghai, and Chongqing, are 0.0302, 0.0357, 0.0271, and 0.0321, respectively. Reunitizing the four numbers yields 0.2416, 0.2851, 0.2167, and 0.2566, which are close to the results shown in Tables 3 and 4.

**Table 7 The GRP-based values of ASI, AASI, and average RRG of Mainland China's 31 regions (1998-2012)**

| Region | ASI ($\hat{w}_i$) | AASI($\hat{w}_i^*$) | RRG | Region | ASI ($\hat{w}_i$) | AASI($\hat{w}_i^*$) | RRG |
|--------|-------------------|---------------------|-----|--------|-------------------|---------------------|-----|
| Anhui | 0.0307 | 0.0307 | 0.1416 | Jiangxi | 0.0326 | 0.0326 | 0.1478 |

| Beijing | 0.0302 | 0.0303 | 0.1417 | Jilin | 0.0331 | 0.0331 | 0.1539 |
|---|---|---|---|---|---|---|---|
| Chongqing | 0.0321 | 0.0321 | 0.1507 | Liaoning | 0.0302 | 0.0302 | 0.1418 |
| Fujian | 0.0301 | 0.0301 | 0.1412 | Ningxia | 0.0361 | 0.0361 | 0.1688 |
| Gansu | 0.0301 | 0.0301 | 0.1396 | Qinghai | 0.0350 | 0.0350 | 0.1661 |
| Guangdong | 0.0306 | 0.0307 | 0.1396 | Shaanxi | 0.0363 | 0.0363 | 0.1712 |
| Guangxi | 0.0330 | 0.0330 | 0.1488 | Shandong | 0.0337 | 0.0338 | 0.1528 |
| Guizhou | 0.0325 | 0.0325 | 0.1557 | Shanghai | 0.0271 | 0.0272 | 0.1234 |
| Hainan | 0.0297 | 0.0297 | 0.1411 | Shanxi | 0.0332 | 0.0332 | 0.1501 |
| Hebei | 0.0311 | 0.0312 | 0.1432 | Sichuan | 0.0322 | 0.0322 | 0.1486 |
| Heilongjiang | 0.0271 | 0.0271 | 0.1264 | Tianjin | 0.0357 | 0.0357 | 0.1694 |
| Henan | 0.0334 | 0.0334 | 0.1507 | Xinjiang | 0.0308 | 0.0308 | 0.1470 |
| Hubei | 0.0325 | 0.0324 | 0.1512 | Xizang (Tibet) | 0.0313 | 0.0313 | 0.1547 |
| Hunan | 0.0329 | 0.0328 | 0.1536 | Yunnan | 0.0285 | 0.0285 | 0.1319 |
| Inner Mongolia | 0.0424 | 0.0425 | 0.1924 | Zhejiang | 0.0320 | 0.0320 | 0.1475 |
| Jiangsu | 0.0338 | 0.0338 | 0.1550 | *Mean* | *0.0323* | *0.0323* | *0.1499* |

**Note**: RRG represents the mean of relative rates of growth of the GRP from 1998 to 2012.

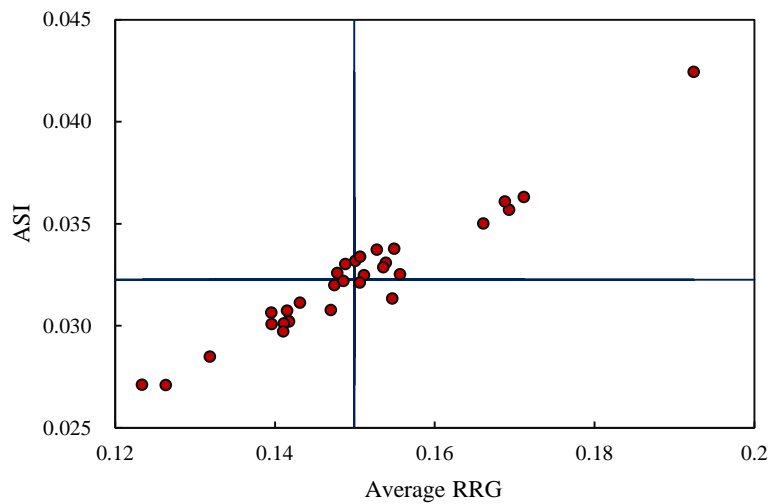

**Figure 5 The relationship between ASI and the average RRG of Chinese GRP (1998 to 2012)**

**Note**: The horizontal line represents the mean of ASI, 0.0323, and the vertical line denotes the mean of the average RRG, 0.1499.

The ASI is derived from the allometric scaling exponents, which is theoretically equal to the ratios of RRG. In a sense, the ASI is equivalent to the relative growth rates. However, there are errors between ASI values and RRG values in an empirical analysis. In this case, we can plot the relationship between ASIs and the means of RRGs to show the growing feature of Chinese regional system (Figure 5). The continuous form of the RRG expression is $r=\mathrm{d}Q(t)/[Q(t)\mathrm{d}t]$ $=\mathrm{d}\ln Q(t)/\mathrm{d}t$. This means that the relative growth rates of $Q(t)$ are equal to the absolute growth rate of $\ln Q(t)$. For simplicity, let time difference $\Delta t=1$. Then, discretizing this expression yields two formulae of RRG such as

$$r = \frac{\Delta Q_t}{Q_t \Delta t} = \frac{Q_t - Q_{t-1}}{Q_{t-1}}, \qquad (29)$$

and

$$r = \frac{\Delta \ln Q_t}{\Delta t} = \ln Q_t - \ln Q_{t-1}, \qquad (30)$$

where $Q_t$ is the discretizing result of $Q(t)$. In theory, equations (29) and (30) are equivalent to one another, but in practice, they are different due to the errors stemming from discretization. If and only if $\Delta t \rightarrow 0$, the results from the two formulae are identical to each other. In this work, equation (29) is employed to estimate the RRG as it is easy to understand.

The difference of allometric growth of Mainland China's 31 regions can be illustrated with the plot (Figure 5). By average values of ASIs and average RRGs, the plot can be divided into 4 quadrants. The upper right quadrant represents the higher allometry, and the lower left quadrant represents the slower allometry. The rectangular coordinate method developed in spatial autocorrelation analysis and principal component analysis can be employed to make a simple cluster analysis. The cluster method based on coordinate systems is simple and clear. In terms of allometric growth and economic development, the 31 regions can be classified as 6 types: (1) the developed region with higher allometric growth, e.g., Tianjin, Jiangsu; (2) the developed regions with lower allometric growth, e.g., Shanghai, Guangdong; (3) the developing/undeveloped regions with higher allometric growth, e.g., Inner Mongolia, Ningxia, Qinghai; (4) the developing/undeveloped regions with lower allometric growth, e.g., Gansu, Yunnan; (5) the developed regions with median allometric growth, e.g., Beijing, Zhejiang; (6) the developing/undeveloped regions with median allometric growth, e.g., Xizang (Tibet), Xinjiang. The allometry types can be determined by the scatterplots such as Figure 5, and the types of economic development can be judged by the data of population size, GRP, and level of urbanization (the original datasets are attached in MMC2 File).

In the similar way, the MAS analysis can be applied to Mainland China's regional system by means of another measurement, level of urbanization. The sample path of the urbanization level is short ($T=9$), but the results can be shown for reference. The scaling consistency ratio is *SCR*=0.0247<0.1, and the ACC $R^*$=0.9556 > $R_c$=0.7977 (significance level $\alpha$=0.01). The ASI values are close to the AASI values. The result can be shown by a histogram (Figure 6). An urbanization curve can be divided into four sections: initial stage, acceleration stage, deceleration stage, and terminal stage (Chen, 2014a). The three municipalities, Beijing, Shanghai, and Tianjin had reached the terminal stage, and their urbanization took on lower allometry. The relative growth rates of the three provinces in Northeast China, Heilongjiang, Jilin, and Niaoning, were also lower from 2005 to 2013. The main regions of higher allometry are the undeveloped provinces and autonomous regions in West China and Southwest China such as Guizhou, Shaanxi, Yunnan, Sichuan, Gansu, and Guangxi. This indicates that Chinese urbanization depends chiefly on preliminary industrialization based on the secondary sector. Industrial transition influences urban evolution and thus influences urban scaling. The regional economies become more complex when they evolve from mining and exploitation of raw resources into manufactures and services (Pumain *et al*, 2006). To some extent, the allometric growth of Chinese urban and regional systems can be understood by industrial and economical evolution.

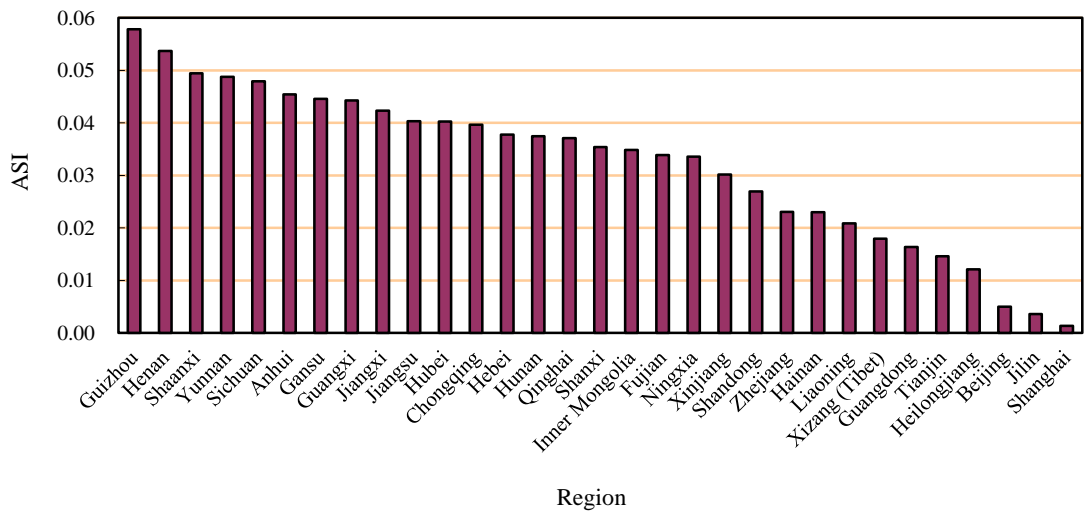

**Figure 6 The ASI values of the level of urbanization of Mainland China's 31 regions (2005-2013)**

The MAS model can be employed to explain spatial heterogeneity or the spatial agglomeration of urban and regional evolution. Suppose there are a group of human settlements of the same size in a region. The settlements form a pattern of spatial homogeneity where the size distribution is concerned. However, the relative rates of growth of these settlements are different owing to spatial difference of geographical conditions. Different settlements have different ASI values. The settlements with higher ASI values become larger and larger over time. Thus, the system of human settlements evolves from homogeneity into heterogeneity: the homogeneous distribution of cities and towns changes to the rank-size distribution. If the ASI value of a city is forever constant, the smaller settlements will have no chance to develop. In fact, the allometric competition is far more brutal than the predator-prey interaction in an ecologic system: the latter often causes periodic oscillation of two correlative species, while the former can bring one of the two species to extinction (Bertalanffy, 1968). Similarly, simple allometry may result in disappearance of smaller towns. Fortunately, the law of allometry is not an iron law of growth for geographical systems. The allometric scaling exponent is not a real constant. It can be treated as a constant in a certain period. When a city becomes bigger and reaches to its capacity of space and size, its ASI can jump from one larger constant to a smaller constant; meanwhile, smaller cities can obtain higher ASI values (Chen, 2008a; Chen, 2014a). Thus the system of settlements will evolve from a heterogeneous pattern into another heterogeneous pattern, or even regress to spatial homogeneity in some local regions. As far as China is concerned, generally speaking, the developed regions such as Shanghai, Beijing, and Guangdong, the old industrial regions such as Heilongjiang, while the undeveloped regions such as Xizang (Tibet) have no significant advantage of allometric growth. An exception is Jiangsu Province. The developed and old industrial regions reached their capacities, and it is not the opportunity for some undeveloped regions. In terms of GRP growth and urbanization, the ascendant areas are mainly the less developed regions with rich mineral resources, including Inner Mongolia, Shaanxi, and Qinghai. This suggests that Chinese pattern of spatial heterogeneity had been changing in the last ten years due to mining and exploitation of raw resources and development of manufactures and services.

## 4 Discussion

A scientific research should proceed first by describing how a system works (by mathematics or measurements) and later by understanding why (by observations, experiences, or experiments) (Gordon, 2005; Henry, 2002). In order to describe a thing exactly in a proper way, we must find its characteristic scale. A characteristic scale corresponds to a 1-dimensional variable in Euclidean geometry, and often termed "characteristic length" (Takayasu, 1990). The traditional mathematical methodology is very successful in describing the simple thing with characteristic lengths. However, complex systems such as cities have no characteristic scale and cannot be effectively described using the conventional mathematical theories. This type of systems belongs to the scale-free phenomena, which take on the property of scale invariance (Chen, 2008a). In this case, we can use scaling method to find a characteristic parameter for quantitative analysis. Scaling indicates invariance of contraction and dilation transform. For a function of variable $x$, $f(x)$, it involves scaling law if and only if it satisfies the relation $Tf(x)=f(\lambda x)=\lambda^b f(x)$, where T denotes contraction/dilation transform, $\lambda$ refers to scale factor, and $b$ to scaling exponent. Fractal geometry is one of the powerful tools of scaling analysis, by which we can find a set of scaling exponents including fractal dimension from scale-free patterns and processes. A scaling exponent has its characteristic scale and can reflect the essential property of a complex system. Generally speaking, a good mathematical model is always a characteristic function (eigen fucntion) of some mathematical transform, and the key parameter is the corresponding characteristic values (eigen values). A fractal model is just the characteristic function of scaling transform, and the fractal dimension or the related scaling exponent is the characteristic value.

However, new problems arise. First, sometimes it is difficult to evaluate a fractal dimension for a complex system. Especially, it is hard to calculate a determinate fractional dimension based on a dynamic process of urban evolution. Second, the values of fractal dimension depend on the measuring methods. Different measurement approaches result in different results of fractal dimension estimation (Batty and Longley, 1994; Frankhauser, 1994; Frankhauser, 1998; Longley and Batty, 1989). Where there is an immensurable quantity, there is a type of symmetry, which is defined as invariance of transformation (Lee, 1988). Just because of the immensurability of coastline length, Mandelbrot (1983) found fractional dimension and put forward the concept of fractals. Now, the immensurability of some fractal dimension maybe suggests a super-symmetric nature of complex systems (Chen, 2008a). This conundrum remains to be solved in future. To a certain extent, the allometric scaling can be employed to analyze the systems without specific fractal dimensions (Chen, 2010). The use of a parameter rests with its relative numerical quantity or the comparable relationship between different parameters rather than its absolute quantity. An allometric scaling exponent concerns the ratio of two fractal dimensions, which is just defined in a comparable framework of different parameters. Based on this idea, the MAS analysis is proposed. The method avoids the direct measurement of fractal dimensions, and makes use of the relative quantities of a set of fractal parameters.

The MSA modeling, algorithms of parameter estimation, statistical tests, and typical example have been shown above. The theoretical basis and the local-local allometric model of MSA have been preliminarily presented years ago (Chen, 2008a; Chen and Jiang, 2009), but the global-local allometric model and the approaches of algorithms and tests have not be developed. The focus of this article is on the expansion of models, the method of calculation, the measurements of statistical test, and the cases of application. The strong points of MSA analysis are obvious. It is simple, readily understandable, and can be used to examine the spatio-temporal evolution of an urban or regional system. Using this methodology, we can reveal the comparative advantage and competitive power of different cities or regions. Where urban research is concerned, the limitations of MSA are as follows. Firstly, the urban system concerned must comply with the law of allometric growth to some extent. Otherwise, the method may be invalid. Secondly, the sample paths of time series must be long enough. Empirical studies show that the sampling results should include at least ten years of data points. It is better if the data series are uninterrupted. Thirdly, the results only reflect the relative level of growth of an urban/regional system. It cannot bring to light the absolute level of urban/regional development. The defects cannot obscure the virtues of the MSA modeling. In fact, every method has its own shortcomings. The precondition of effective application of a method is to learn its sphere of application. Despite all the problems mentioned above, it has a potential to improve the current approaches to spatial analysis of human geography in the perspective of nonlinear dynamics. In principle, allometry indicates that two dynamic functions are directly linked, but not all types of functions can generate dynamics. The allometric scaling relation based on logistic growth suggests spatial replacement dynamics of urban evolution (Chen, 2014a). If the observational data are reliable, the MSA analysis will provide useful information of a geographical system.

The MSA is associated with the classification of geographical space. By different allometric scaling relations, geographical space can be divided into three types: real space (R-space), phase space (P-space), and order space (O-space) (Chen, 2012a; Chen, 2014b). The real space is defined for spatial patterns, and can be described with spatial datasets based on maps and remote sensing images. The phase space is defined for dynamic process, and can be described with time series data. The order space is defined for cascade structure, and can be described with cross-sectional data. Compared with Euclidean geometry, fractal geometry goes beyond the limit of spatial form and can be employed to model function and information of a complex system. Using fractal dimension, we can describe the real space, phase space and order space of geographical systems. For a city, urban form can be examined in the real space, urban growth can be reflected in the phase space, and urban internal structure can be realized through the order space. The three-type space theory can be used to make clear many questions which are confused and puzzle geographers for a long time. For example, the parameter relationship between Zipf's law and the allometric scaling law used to cause misunderstanding. Today, this problem can be easily solved using the concepts from the new space theory. In fact, where order space is concerned, an allometric scaling relation can be derived from a pair of Zipf's distributions. Consider a region with $n$ cities which follow Zipf's law. A Zipf distribution can be transformed into a hierarchy with cascade structure (Chen, 2012b; Chen, 2014b), which is equivalent to a network structure (Batty and Longley, 1994). Suppose that city sizes are measured with urban population, and the fractal dimension of the $k$ city's population distribution is $D_{(k)}$ (a local parameter). A parameter relation can be derived as follows (Chen, 2014b)

$$q = \frac{D_p}{D_n} = \frac{1}{nD_n} \sum_{k=1}^{n} D_{(k)} \, , \qquad (31)$$

where $k$ is the rank of a city ($k$=1,2,…, $n$), $q$ denotes the Zipf's exponent, $D_n$ refers to the fractal dimension of the network of the $n$ cities (a global parameter), and $D_p$ to the average fractal dimension value of the $n$ cities' population, that is

$$D_p = \frac{1}{n} \sum_{k=1}^{n} D_{(k)} \, . \qquad (32)$$

According to equation (13), the allometric scaling exponent is $\alpha_{ij} = D_{(i)}/D_{(j)}$, where $i$, $j$=$k$=1,2,…, $n$. If and only if $i$=$j$, we have $\alpha_{ij}$=1, or else $\alpha_{ij} \neq 1$. On the other hand, if and only if $D_n$=$D_p$, then $q$=1, otherwise $q \neq 1$. Whether or not $\alpha_{ij}$=1 has no relation with whether or not $q$=1.

The main limitations of this study rest with two aspects. One is absence of efficient spatial representation and display of analytical process and results. In fact, the MSA method can be integrated into the technology of geographical information system (GIS). If so, the spatio-temporal evolution of regional and urban structure can be visually illustrated in the right perspective. The other is related to data quality and sample sizes. The sample paths are not long, and the materials are statistical data based on top-down abstraction rather than what is called "big data" based on bottom-up production. Despite all that, as a methodological study, this paper expands the concepts, models, analytical approaches, and explanatory power of allometric scaling analysis.

Recent years, an increasing number of scholars and researchers have realized the importance of the scaling analysis in urban studies. However, a large amount of research of uneven in quality leads to confusion of scaling concept. Despite many exciting achievements, a number of basic questions have not been answered. *First, real scaling and false scaling*. As indicated above, a city has its two aspects. One is characteristic scale, and the other is scale free. The former can be described with traditional mathematical methods and conventional measurements, and the latter should be characterized with scaling exponents. A scaling analysis can be made on the scale-free aspects rather than the aspects with characteristic scales (Chen, 2008a). However, scaling and characteristic scale of cities are sometimes confused with one another in literature. *Second, scaling range*. A scaling relation is valid and stable within certain range of scales. If the scale is too large or too small, the relation may break down (Bak, 1996; Chen, 2008a; Chen, 2012b). Sometimes, scaling range is not clear, but the size threshold influences the estimation of allometric exponents. A recent discovery is that the scaling exponent values of the allometric relation between patents and city sizes depend on the population size cut-offs (Arcaute *et al*, 2015). *Third, spatial measurements*. Urban boundaries impact scaling relation and exponent. The allometric scaling relation between urban area and population depends on spatial definitions of the city. This allometry can be revealed on the base of urbanized area (UA), but cannot be brought to light by city proper (CP) and metropolitan area (MA) (Chen, 2008b). A meaningful discovery is that the scaling exponent values of the allometric relation between urban $CO_2$ emissions and city population sizes depend on the definition of urban area (Louf and Barthelemy, 2014a; Louf and Barthelemy, 2014b). *Fourth, exponential allometry and logistic allometry*. General allometric scaling is based on a pair of processes of exponential growth or dual patterns of exponential distributions (Bertalanffy, 1968; Chen, 2008a). However, if we investigate a set of long sample paths of time series, the explicit allometric scaling based on exponential growth may change to a latent allometric scaling based on logistic growth (Chen, 2014a). In this instance, the simple power-law allometry is disabled and cannot yield acceptable parameter values. *Fifth, allometric scaling degeneration.* Differing from the natural laws in classical physics, urban laws are not of translational symmetry in space and time. Instead, urban laws are of scaling symmetry. Thus, the mathematical structure and parameter values of urban laws are not determinate (Chen, 2008a). Allometric scaling laws are essentially fractal relations (Chen and Liu, 1998). The fractal structure and allometric scaling can be disturbed by governmental actions (Chen, 2014a). As a result, a power law relation may degenerate to exponential relation, logarithmic relation, or even linear relation (Chen, 1995; Chen and Wang, 1997). In this case, the scaling exponents cannot be used to effectively characterize urban systems. By dimensional analysis, the reasonable scaling exponent of urban area and population comes between 2/3 and 1 (Lee, 1989). The expected value is about 0.85 (Chen, 2008a; Chen, 2010), and the empirical values are close to 0.85 (Louf and Barthelemy, 2014a; Chen, 2010). However, it may exceed 1 if a city is wrongly managed and planned (Chen, 2008a). Owing to all that, urban scaling is often misunderstood by a few students. It is hard to clarify all the related questions in a few lines of words. The pending problems remain to be explored and solved in future.

## 5 Conclusions

The main academic contributions of this study rest with three aspects. In theory, it develops a global-local allometric model to complement the local-local allometric analysis, and especially, reveals the mathematical relationships between scaling exponents, growth rates, and standard derivations. In practice, it demonstrates the equivalence relation between element-element allometry and the element-system allometry. In methodology, it presents a complete analytical process including models, algorithms, statistical tests, and typical examples. In particular, a new algorithm and two statistic test method are proposed. From the theoretical derivation and empirical analyses, four main conclusions can be drawn as follows. **First, the MSA analysis can be applied to the comprehensive evaluation of the relative levels of urban growth and regional development.** The MSA is based on FPM, and the fractal dimension ratio of two measures equals the ratio of relative growth rates of two corresponding elements, which in turn equals the ratio of the standard deviations of the two logarithmic variables. Thus the MSA method can be employed to compare the developing potentials and predict the growing trends of different elements of an urban/regional system. **Second, the ASI can reflect the ratio of the relative growth rate of an element to that of the system.** To some extent, the ASI values are based on the part-part scaling, but they approximate to the part-whole scaling exponents. In other words, the scaling index of the allometric relations between an element (part) and all other elements (parts) is close to the scaling exponent of the allometric relation between an element and the system (whole). The part-whole scaling exponent indicates the ratio of the relative growth rate of an element to that of the system. **Third, there are different equivalent approaches to evaluating the ASIs.** Two methods can be adopted to generate the SEM: one is the least squares method, and the other, the standard deviation method. The latter suggests a theoretical relationship between characteristic scales (mean, standard deviation) and scaling (allometric exponent, fractal dimension). At least three approaches (matrix power, arithmetical averaging, geometric averaging) can be applied to the estimation of the ASIs, and among these methods, the simplest one is the geometric averaging method. **Fourth, two measurements can be used to make statistical tests: one is the scaling consistency index, and the other is the mean of the correlation coefficients**. The index of scaling consistency is defined by analogy with the principle of AHP, while the ACC is based on the LSR method, by which we can get a matrix of fractal parameters. There is no perfect method for the testing of a modeling result. All the processes and effects of statistical tests are for reference only. The quality of a mathematical model or method is finally judged and evaluated by its effect of application to real natural and social systems.

## E-Components (Online Supplementary Material) Legends

**MMC1 File. The demonstration of the relation between allometric scaling exponents and the ratios of standard deviations (DOCX)**.

In this Word file, the following equivalence relation is derived: if the relation between $x$ and $y$ follows the allometric scaling law, the scaling exponent $b$ will equal the ratio of the standard deviation of $\ln(y)$ to the standard deviation of $\ln(x)$. The scaling exponent is a parameter for scale-free analysis, while the standard derivation is a characteristic scale for statistic description. This finding suggests the unity of opposites between characteristic scales and scaling exponents.

**MMC2 File. The datasets of GRP, level of urbanization, and population for 31 Chinese regions and a simple calculation example (1998-2012) (XLSX).**

In this Excel file, the panel datasets of gross regional product (GRP), level of urbanization, and population at year-end are shown for the 31 regions of mainland China. Based on the four municipalities directly under the Central Government of China, an Excel-based complete process of multiscaling allometric analysis is provided in an understandable way. Using these datasets, readers can testify the analytical processes; imitating the simple case, readers can make urban allometric scaling analysis.

**MMC3 File. Two Matlab programs for calculating element-element allometric scaling exponents (MAT).**

In this m-file, two complete Matlab-based calculation programs are provided for readers to compute urban element-element (part-part) allometric scaling exponents. One program is written by means of the least squares method, and the other one is composed by way of the equivalence relation between allometric scaling exponents and the ratios of two standard deviations of logarithmic variables (see MMC1 File). Using any one of the two programs, readers can make multi-scaling allometric analyses of cities by substituting new data for the given data (see MMC2 File for the given data).

**MMC4 File. A Matlab program for calculating element-system allometric scaling exponents (MAT).**

In this m-file, a simple Matlab-based program is provided for readers to compute the element-system (part-whole) allometric scaling exponents. Using this program, readers can calculate the approximate scaling exponents of the urban systems following the law of allometric growth.

## Acknowledgements

This research was sponsored by the National Natural Science Foundations of China (Grant No.

41590843 & 41171129). Many thanks to the four anonymous reviewers, whose interesting constructive comments are helpful for improving the quality of this paper.